\documentclass[letterpaper, 10 pt, conference]{ieeeconf}  

\IEEEoverridecommandlockouts                              

\usepackage{graphicx}

\usepackage{amsmath} 
\usepackage{amssymb} 

\title{\LARGE \bf
Rendering Forces With a Modular Cable System, Motors, and Brakes}

\author{Jan Ulrich Bartels$^{1,2,*}$, Alexander Achberger$^{2,*}$, Katherine J. Kuchenbecker$^{1,\dagger}$ and Michael Sedlmair$^{2,\dagger}$
\thanks{$^{1}$Haptic Intelligence Department, Max-Planck Institute for Intelligent Systems, Heisenbergstraße 3, 70569 Stuttgart, Germany.
        {jub@is.mpg.de, kjk@is.mpg.de}}%
\thanks{$^{2}$Visualisierungsinstitut (VISUS), Universität Stuttgart, Allmandring 19, 70569 Stuttgart, Germany.
        {alexander.achberger@vis.uni-stuttgart.de, michael.sedlmair@vis.uni-stuttgart.de}}%
\thanks{$^*$equal contribution (co-first). $^\dagger$equal contribution (co-last).}
}
\begin{document}

\maketitle
\thispagestyle{empty}
\pagestyle{empty}

\begin{abstract}

We describe the hardware design, force-rendering approach, and evaluation of a new reconfigurable haptic interface consisting of a network of hybrid motor-brake actuation modules that apply forces via cables.
Each module contains both a motor and a brake, enabling it to smoothly render active forces up to 6\,N using its motor and collision forces up to 186\,N using its passive one-way brake. The modular design, meanwhile, allows the system to deliver rich haptic feedback in a flexible number of DoF and widely ranging configurations.

\end{abstract}
\section{Introduction}

Cable-based robotic systems have a long history in human-computer interaction, beginning with early systems such as SPIDAR~\cite{hirata19923}, introduced in 1992, and subsequent designs that focused on increasing the number of DoF~\cite{lindemann1989construction,williams1998cable,kim2002tension} or enlarging the workspace~\cite{chen2012ifeel6, tarrin2003stringed}. However, deployment of these devices remains rare. We believe one main reason for their low adoption rate is that existing systems have a fixed number of actuators in a fixed configuration and thus cannot be adapted to many specific use cases.

Recent work by Achberger et al.~\cite{achberger2021strive} demonstrated that a modular cable-based design can address several of these limitations. Their STRIVE system allows one to mount a variable number of brakes to static objects and the user's body, thereby adapting the system's DoF and configuration to specific use cases. Additionally, the compact size of the actuation unit and the lack of scaffolding make the system easy to transport and wear. 

However, STRIVE utilizes a locking brake mechanism and can render only two impedances -- free space and hard collision. This binary force output precludes rendering of many common physical properties, such as texture, viscosity, compliance, and weight, limiting STRIVE's ability to produce \textit{expressive} haptic feedback.

To increase the potential of cable-based robotic systems for human-computer interaction, we extended the mobile and reconfigurable design of STRIVE~\cite{achberger2021strive} with hybrid motor-brake actuation to render \textit{both nuanced and strong force feedback} while still supporting a \textit{flexible number and arrangement of actuation modules}.

\begin{figure}
    \centering
    \includegraphics[width=\linewidth]{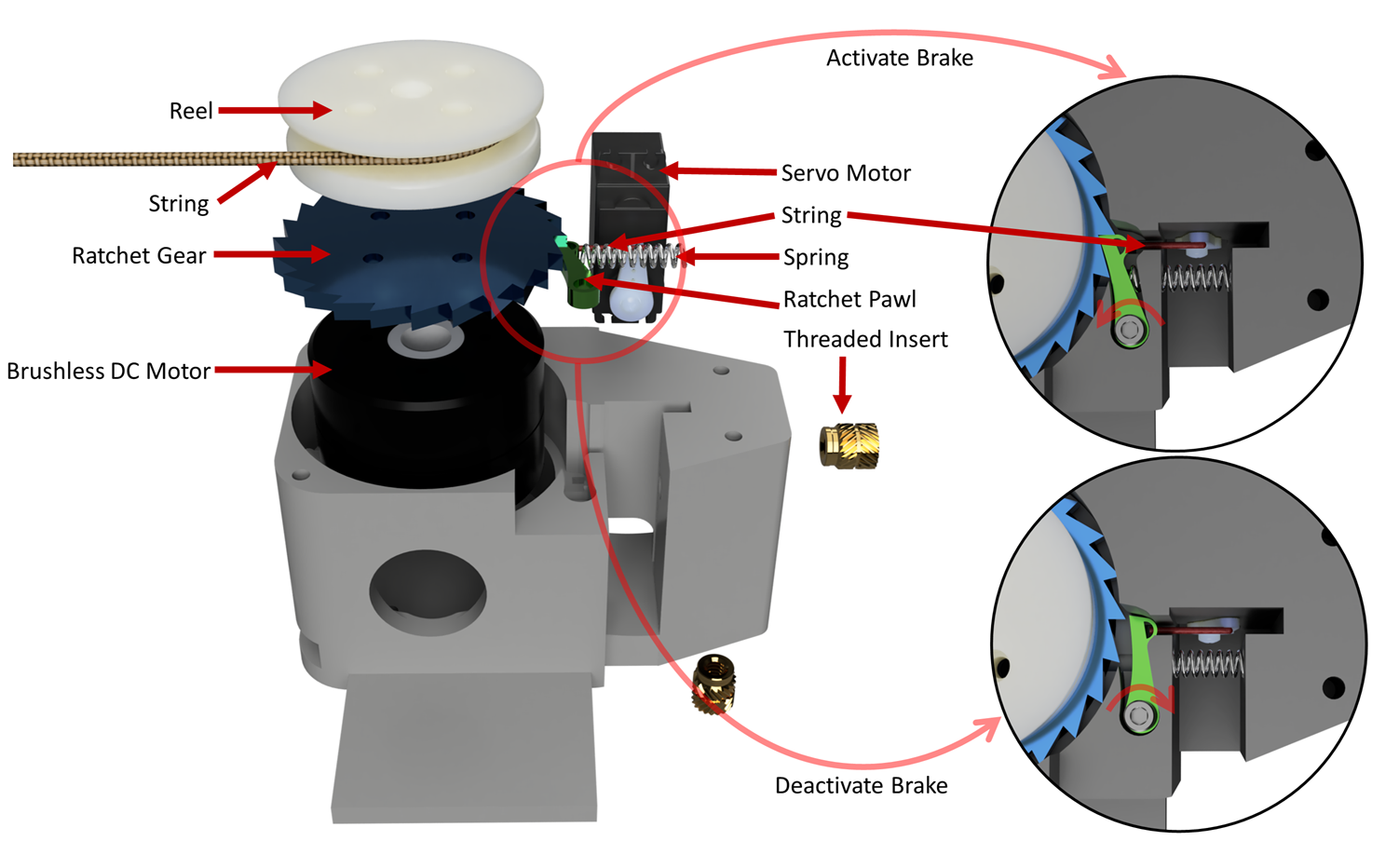}
    \caption{Illustration of the hybrid motor-brake actuation in each module.}
    \label{fig:device}
\end{figure}

\section{Hardware}

\label{sec:hardware}

We selected the BLDC GM3506 gimbal motor (Fig.~\ref{fig:device}, left) to generate active forces and a DS2312 servo motor to actuate the brake. The BLDC motor is driven by the SimpleFOC v2.1 motor driver board~\cite{simplefoc2022}, with its angle measured using an AS5048A magnetic angle sensor.

The motor and brake are controlled by an ESP32 microcontroller, which receives commands via Bluetooth Classic. 
A single module can be built using off-the-shelf components for less than 100\,€, comparable to other accessible haptic devices like Hapkit~\cite{morimoto2014d81} and Snaptics~\cite{zook2021snaptics}.

\section{Software}

The BLDC motor is controlled via the FOC current-control mode of the SimpleFOC library. The motor angle is used to calculate phase currents for torque generation, which are regulated via current sensors and a PID controller. 

\subsection{Modular Force-Rendering Algorithm}
\label{sec:force_calc_method}

A key feature of our approach is the ability to combine multiple modules into a unified force-rendering system that can render up to three DoF of force feedback. The system supports an arbitrary number of devices at arbitrary positions, with the goal of rendering a desired 3D force vector at any target location. Achieving this goal requires computing the cable tensions, which must remain within each module's physical limits, as cables can only pull. In our setup, forces range from 0.5\,N, to keep the cables taut, up to 6.0\,N, corresponding to the motor's maximum steady-state current.

Depending on the number of modules, the system can be \emph{underactuated} or \emph{redundantly actuated}. With more than four devices properly distributed, the system is redundantly actuated, providing an infinite number of solutions for all force directions. With fewer devices, an exact solution will exist only if the desired force vector lies within the space spanned by the available force vectors from each device; otherwise, the nearest reachable force is rendered.
The objective is therefore to determine feasible tensions for each device that produce the desired force vector while respecting their physical capabilities.

This problem corresponds to the classical bounded-tension issue in cable-driven parallel manipulators, which has been extensively analyzed, e.g., by Pott and Bruckmann~\cite{pott2013cable}.
We adopt the approach of Hassan and Khajepour~\cite{hassan2011analysis}, who use Dykstra's algorithm to separate the cable tensions into two components: one that produces the required force at the end effector, and another that lies in the system's null space, allowing internal tensions to be adjusted without affecting the output force. 
Feasible tension solutions are found by projecting onto the intersection of the set of allowable tensions and the equilibrium subspace, ensuring all physical limits are respected.
We extended the algorithm with a specific starting point in the solution space to find solutions with minimum tensions, reducing energy consumption and motor heating.

A key advantage of Dykstra's method is its robustness: even with too few modules or an infeasible force direction, it still returns the tension vector with the smallest Euclidean distance to feasibility. This approach makes the method stable and well-suited to real-time active force rendering in our system.

\section{Technical Validation}

To test 3-DoF force rendering, we set up four modules connected to an ATI Nano17 SI-16-0.1 force sensor sampled at 1000\,Hz via a National Instruments USB-6366 USB X DAQ. 
Three modules were arranged in a roughly equilateral triangle, with the sensor located roughly at the center of the triangle, 30\,cm above the plane. A fourth module was located about 2\,m directly above the sensor.

We rendered 182 force vectors distributed over a sphere with a radius of 1.5\,N, corresponding to the tension from a single motor at 0.5\,A. 
The force sensor was zeroed at the start with all motors applying zero tension. Force vectors were then commanded sequentially, each held for 1\,s, with measurements averaged.

\begin{figure}
    \includegraphics[width=0.49\linewidth]{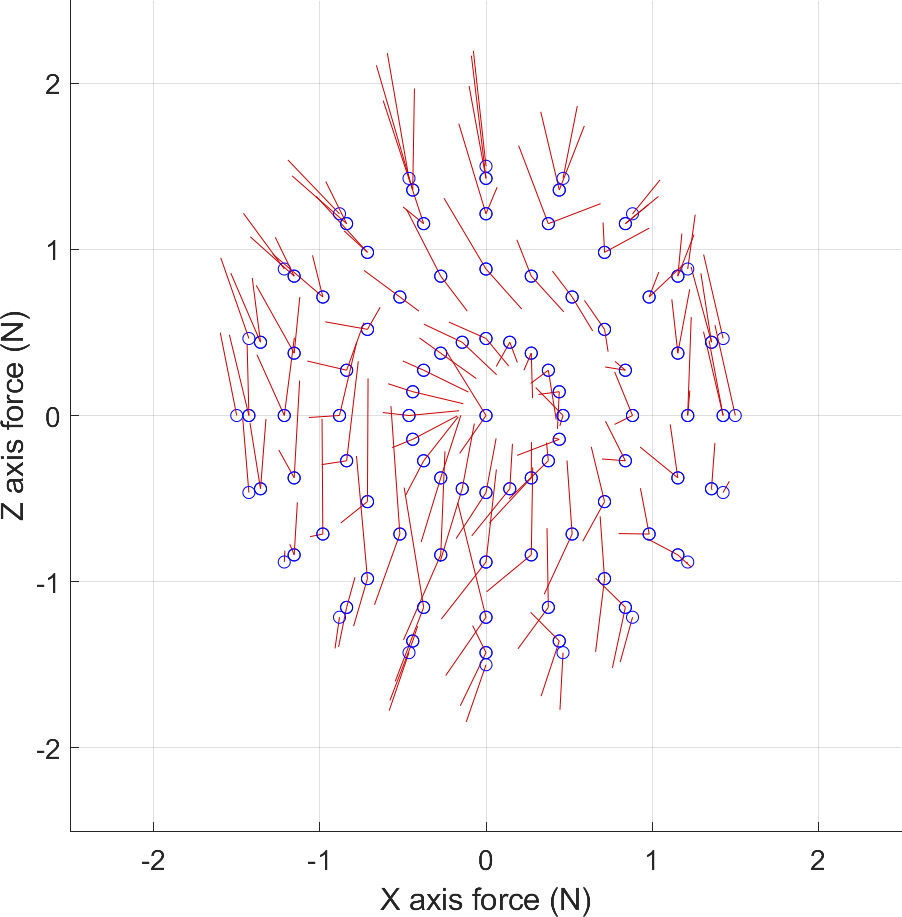}
    \includegraphics[width=0.49\linewidth]{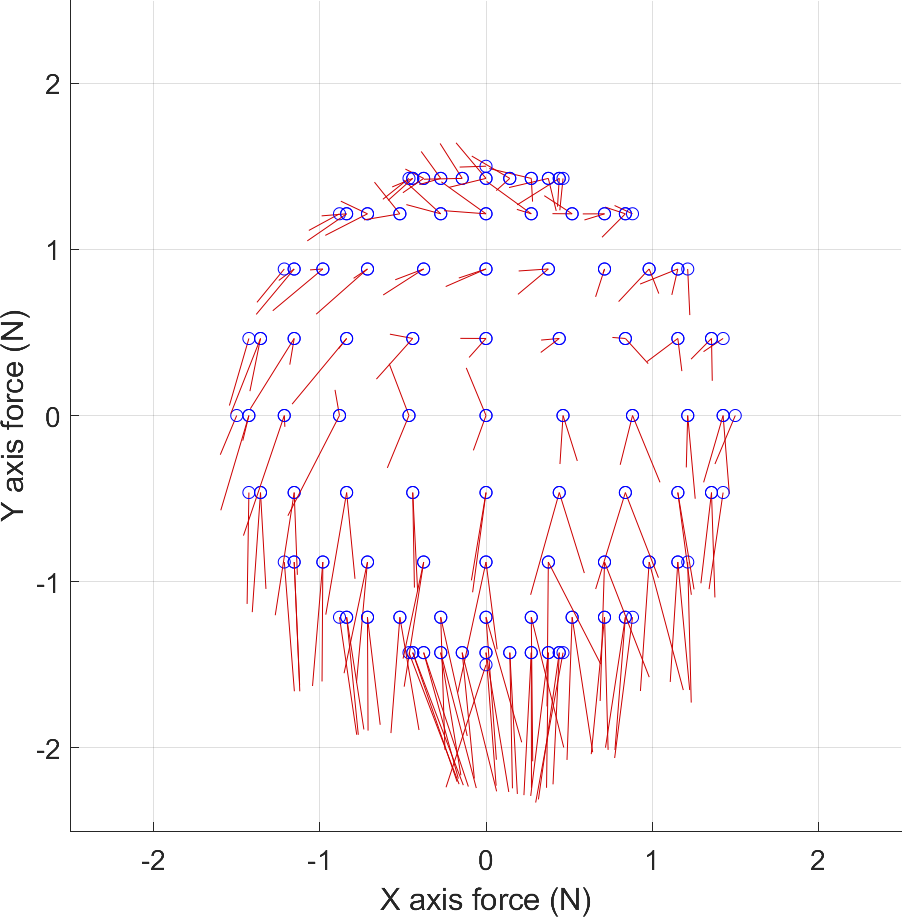}
    \caption{Intended vs. measured force. Blue circles denote the intended force vectors, and red vectors show the corresponding errors of the rendered forces.} 
    \label{fig:Eval_3D_Setup_A}
\end{figure}

Since the force sensor was manually oriented, a small unknown rotation existed between its reference frame and the devices' coordinate system. We estimated this rotation by rendering three basis vectors, aligning the sensor's Z-axis with the devices' Y-axis, and determining the Z-rotation to minimize the errors between the desired and measured basis vectors. 
The results shown in Fig.~\ref{fig:Eval_3D_Setup_A} illustrate how the measured force vectors corresponded to the commanded vectors across the workspace.

The rendered forces generally align with the intended vectors with an average angle error of 14.0$^{\circ}$. Importantly, all measured force vectors fell within 45$^{\circ}$ of their intended directions, a threshold identified by Reel Feel~\cite{devrio2025reel} as sufficient for convincing force feedback in virtual reality (VR). The measured magnitudes also closely matched expectations, averaging 1.84\,N  compared to the commanded 1.5\,N, with an average magnitude error of 0.58\,N.

\section{User Study}

\begin{figure}[t]
    \centering
    \includegraphics[width=\linewidth]{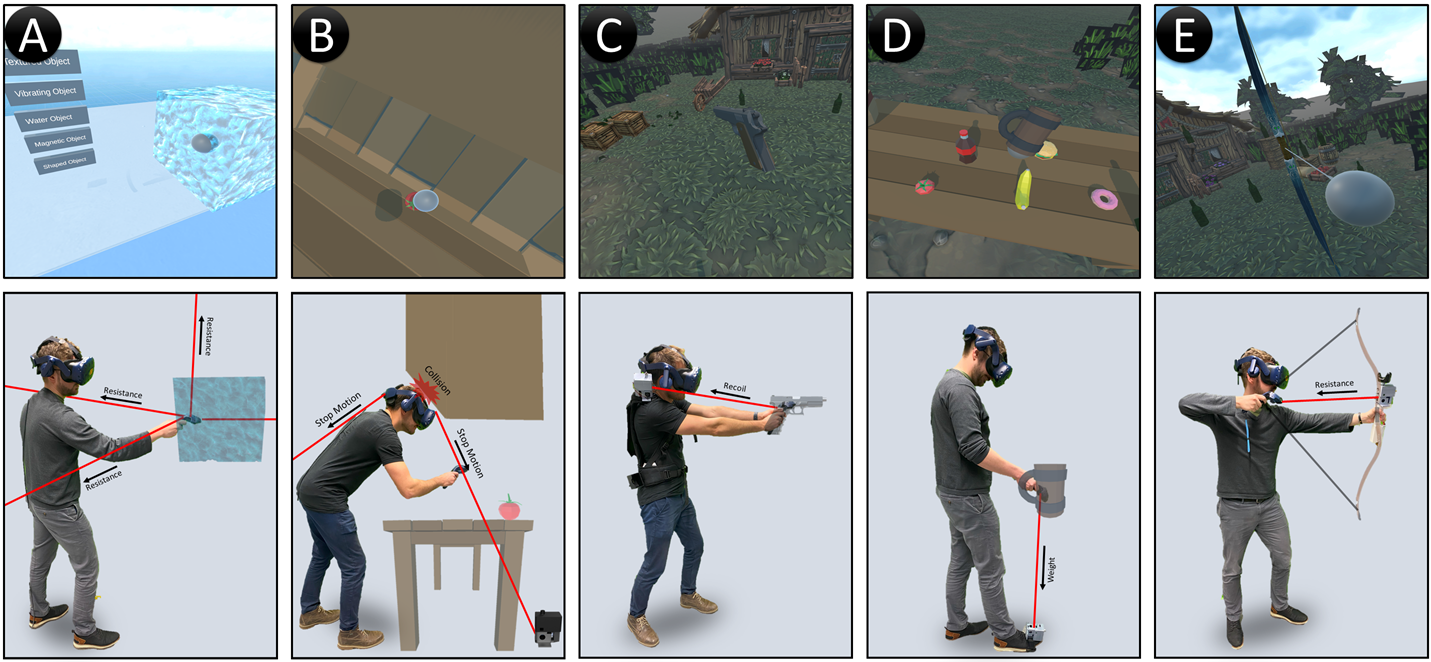}
    \caption{The five VR scenarios featured across our user studies. On top: the view in VR. On the bottom: the system configuration. The five scenarios are: (A) material simulation, (B) body collision, (C) recoil simulation, (D) weight simulation, and (E) bow simulation.}
    \label{fig:applications}
\end{figure}

We conducted a user study with the five different VR applications shown in Fig.~\ref{fig:applications}.
The first application showcases grounded 3-DoF haptic rendering by allowing users to interact with five virtual materials: magnetic, textured, shaped, vibrating, and water, each defined by distinct combinations of attraction, stiffness, vibration, friction, and damping behaviors.
In the other applications, we further demonstrate our system's versatility by rendering head collisions, firearm recoil, object weight, and bow tension through different body- and static-mounted configurations and force profiles.
Study results show that the haptic feedback strongly enhanced fun, immersion, consistency, and saliency; haptic realism was rated slightly lower, mainly due to occasional mismatches between expected and rendered force directions and timing. Despite these limitations, several applications, especially magnetic materials, head collisions, and body-mounted interactions, were perceived as highly realistic, and overall scores were comparable to those of specialized haptic devices, supporting our goal: the presented modular cable robot uses motors and brakes to effectively deliver expressive haptic feedback across diverse VR applications.

\section*{Acknowledgments}
We thank IMPRS-IS for supporting Jan U.\ Bartels. Supported by DFG (German Research Foundation) under Germany's Excellence Strategy – EXC 2120/1 – 390831618.

\bibliographystyle{ieeetr}%
\bibliography{sample-base}

\end{document}